\documentclass[10pt,final,doublecolumn]{IEEEtran}
\usepackage{amsmath}
\usepackage{amssymb}
\usepackage{amsfonts}
\usepackage{latexsym}
\usepackage{graphicx}
\usepackage{bbding}
\usepackage{enumerate}
\usepackage{subeqnarray}
\usepackage{multicol}
\usepackage{color}
\usepackage{slashbox}
\usepackage{setspace}
\usepackage{mathrsfs}
\usepackage{algorithm,algorithmic}
\usepackage{bm}
\IEEEoverridecommandlockouts
\allowdisplaybreaks[4]

\begin{document}

\title{Massive Access for 5G and Beyond}
\author{
Xiaoming Chen, \IEEEmembership{Senior Member, IEEE}, Derrick Wing Kwan Ng, \IEEEmembership{Senior Member, IEEE}, Wei Yu, \IEEEmembership{Fellow, IEEE}, Erik G. Larsson, \IEEEmembership{Fellow, IEEE}, Naofal Al-Dhahir, \IEEEmembership{Fellow, IEEE}, and Robert Schober, \IEEEmembership{Fellow, IEEE}
\thanks{X. Chen is with the College of Information Science and Electronic Engineering, Zhejiang University, Hangzhou 310027, China (e-mail: {\tt chen\_xiaoming@zju.edu.cn}).}
\thanks{D. W. K. Ng is with the School of Electrical Engineering and Telecommunications, the University of New South Wales, NSW 2052, Australia (e-mail: {\tt w.k.ng@unsw.edu.au}).}
\thanks{W. Yu is with the Department of Electrical and Computer Engineering, University of Toronto, Toronto M5S3G4, Canada (e-mail: {\tt weiyu@comm.utoronto.ca}).}
\thanks{E. G. Larsson is with the Link\"oping University, Dept. of Electrical Engineering (ISY), 58183 Link\"oping, Sweden (e-mail: {\tt erik.g.larsson@liu.se}).}
\thanks{N. Al-Dhahir is with the Department of Electrical and Computer Engineering, the University of Texas at Dallas, TX 75083-0688, USA  (e-mail: {\tt aldhahir@utdallas.edu}).}
\thanks{R. Schober is with the Institute for Digital Communications, Friedrich-Alexander-University Erlangen-N\"{u}rnberg, 91058 Erlangen, Germany (e-mail: {\tt robert.schober@fau.de}).}}\maketitle

\begin{abstract}
Massive access, also known as massive connectivity or massive machine-type communication (mMTC), is one of the main use cases of the fifth-generation (5G) and beyond 5G (B5G) wireless networks. A typical application of massive access is the cellular Internet of Things (IoT). Different from conventional human-type communication, massive access aims at realizing efficient and reliable communications for a massive number of IoT devices. Hence, the main characteristics of massive access include low power, massive connectivity, and broad coverage, which require new concepts, theories, and paradigms for the design of next-generation cellular networks. This paper presents a comprehensive survey of massive access design for B5G wireless networks. Specifically, we provide a detailed review of massive access from the perspectives of theory, protocols, techniques, coverage, energy, and security. Furthermore, several future research directions and challenges are identified.
\end{abstract}


\section{Introduction}
The widespread applications of the Internet of Things (IoT) in a variety of fields, e.g.  industry, agriculture, medicine, and traffic, have spurred an explosive growth in the number of IoT devices \cite{IoT1}-\cite{IoT3}. As of 2017, there were $8.4$ billion connected devices across the world. It has been predicted that this number will surpass 75.4 billion by 2025 \cite{Num1}, \cite{Num2}. This growth rate is tremendous and will further increase over the next decade. It is also believed that the number of IoT devices will eventually reach hundreds of billions with a connection density of 10 million devices per km$^2$ by 2030. This trend acts as the catalyst for speeding up the evolution of IoT to the Internet-of-Everything (IoE).

To allow IoT devices to connect, interact, and exchange data anywhere and anytime, they have to be interconnected wirelessly \cite{IoT4}. Hence, wireless access technology providing reliable communications is the key to unleash the potential of the massive IoT. Currently, IoT devices access various wireless networks mainly via low-cost commercial technologies such as Zigbee \cite{Zigbee}, Bluetooth \cite{Bluetooth}, and WiFi \cite{WiFi}. However, these technologies only support short-range wireless access for a moderate number of devices, e.g.  a few hundred devices in an indoor environment or in a small area. Newly emerging services require IoT devices to have seamless access over a wider range. Thus, the existing technologies can only be adopted as an intermediate solution for serving a small number of IoT devices, but eventually will become a bottleneck for providing reliable wireless access to a massive number of IoT devices. On the other hand, a ubiquitous wireless infrastructure is a key enabler for realizing wide coverage for the IoT. Currently, long range radio (LoRa) and cellular IoT are two main access technologies for low power wide area networks (LPWAN) \cite{LPWAN1}-\cite{LPWAN3}. Compared to the LoRa technology, cellular IoT is more beneficial and economical for service providers as it reuses existing cellular infrastructure. In order to support massive access with a connection density of 1 million devices per km$^2$ with cellular networks, the 3rd generation partnership project (3GPP) has selected massive machine-type communications (mMTC) as one of three main use cases of 5G wireless networks and provided a dedicated specification for cellular IoT in Release 13 in 2015 \cite{3GPP}. In this specification, cellular IoT is categorized as narrow-band IoT (NB-IoT) for fixed and low-rate scenarios and LTE-machine (LTE-M) for mobile and high-rate scenarios \cite{5G1}. Hence, the existing cellular network architecture and technology can serve as a solid foundation for enabling massive IoT in practice. A comparison of existing wireless systems supporting IoT is provided in Table \ref{Tab1}.

\newcommand{\tabincell}[2]{\begin{tabular}{@{}#1@{}}#2\end{tabular}}
\begin{table}\centering
\caption{Comparison of existing wireless systems supporting IoT \cite{5G1}.}
\label{Tab1}
\scriptsize
\begin{tabular}{|c|c|c|c|c|c|}\hline
  & Zigbee & Bluetooth & WiFi & LoRa & Cellular \\
\hline
\hline Spectrum & \tabincell{l}{Unlicensed} & \tabincell{l}{Unlicensed} & \tabincell{l}{Unlicensed} & \tabincell{l}{Unlicensed} & \tabincell{l}{Licensed}\\
\hline Connectivity & \tabincell{l}{Moderate} & \tabincell{l}{Small} & \tabincell{l}{Large} & \tabincell{l}{Massive} & \tabincell{l}{Massive}\\
\hline Throughput & \tabincell{l}{ Moderate}  & \tabincell{l} {Low} & \tabincell{l} {High} & \tabincell{l} {High} & \tabincell{l} {High}\\
\hline Range & \tabincell{l} {Short} & \tabincell{l} {Short} & \tabincell{l} {Moderate} & \tabincell{l} {Long} & \tabincell{l} {Long}\\
\hline Security & \tabincell{l} {Moderate} & \tabincell{l} {Low} & \tabincell{l} {Moderate} & \tabincell{l} {High} & \tabincell{l} {High}\\
\hline Power & \tabincell{l} {Low} & \tabincell{l} {Low}  & \tabincell{l} {High} & \tabincell{l} {Low} & \tabincell{l} {Low}\\
\hline Mobility & \tabincell{l} {No} & \tabincell{l} {No}  & \tabincell{l} {No}  & \tabincell{l} {Yes} & \tabincell{l} {Yes}\\
\hline Latency & \tabincell{l} {Low} & \tabincell{l} {Low}  & \tabincell{l} {Low} & \tabincell{l} {Low} & \tabincell{l} {Low}\\
\hline
\end{tabular}
\end{table}

The key to supporting massive IoT in cellular networks lies in designing appropriate multiple access techniques. In fact, enabling multiple access with limited system resources is an inherent issue in cellular networks. Previous and current cellular networks have employed a variety of effective multiple access techniques, such as frequency division multiple access (FDMA) in the first-generation (1G) wireless networks, time division multiple access (TDMA) in 2G, code division multiple access (CDMA) in 3G, and orthogonal frequency division multiple access (OFDMA) in 4G and 5G \cite{Multipleaccess}. However, it is not a trivial task to realize massive access in B5G wireless networks. First of all, there is a lack of information theoretic concepts for the design of massive access. In particular, conventional information theory commonly focuses on multiple access scenarios with only a small number of devices \cite{IT1}, \cite{IT2}. It is not straightforward to extend the conventional multiple access theory to massive access. In particular, short packets are usually employed in massive access for reducing access latency and decoding complexity at the receivers, which requires a much more sophisticated multiple access theory \cite{Shortpacket1}. Secondly, the commonly adopted grant-based random access protocols may lead to exceedingly long scheduling delays and large signaling overheads \cite{RA1}, \cite{RA2}. In fact, for grant-based random access protocols, each device would have to choose a preamble from a pool of orthogonal sequences for accessing the wireless network. Due to the limited coherence time and sequence length, the number of orthogonal sequences is finite. As a result, in the context of massive IoT, two or more devices would choose the same sequence with a high probability, leading to collisions and failure of wireless access. More importantly, the access delay inevitably increases as the number of devices increases. Thirdly, most existing IoT networks adopt orthogonal multiple access (OMA) techniques \cite{Accesstechnique1}. For example, NB-IoT employs single carrier frequency division multiple access (SC-FDMA) for the uplink and OFDMA for the downlink. Although OMA simplifies the transceiver design, it leads to a low spectral efficiency in general \cite{Accesstechnique2}. In the context of massive IoT, applying OMA over limited radio spectrum is challenging due to the resulting underutilization of the system resources. Fourthly, coverage is a critical issue for low power IoT devices. In order to prolong the battery life of IoT devices, their transmit powers are usually very small, e.g. 23 dBm for NB-IoT \cite{Broadcoverage1}. As a result, the received signal is generally weak for signal detection if the distance between the base transceiver station (BTS) and the device is large. As a remedy, NB-IoT enhances the coverage by adopting re-transmission (i.e., time diversity) and low-order modulation (BPSK/QPSK). These techniques enhance the service coverage at the cost of inefficient utilization of the system resources. In other words, if the number of IoT devices is large, the limited system resources may be insufficient for wide coverage. Fifthly, security issues of massive access have to be investigated carefully. Due to the broadcast nature of wireless channels, confidential wireless signals may also be received by unintended devices, resulting in potential information leakage \cite{Security1}, \cite{Security2}. Traditionally, cryptography-based encryption techniques are employed to guarantee security for wireless access. However, due to the fast evolution of eavesdropping techniques in recent years, providing secure encryption has become much more challenging. Unfortunately, most IoT devices have limited computational capability, such that they cannot utilize sophisticated encryption techniques. Moreover, the limited energy supply of massive IoT is a challenging issue. Currently, most IoT devices are battery-powered with small energy storage capacity. Thus, it is necessary to replace the battery frequently to extend the lifetime of the communication nodes. However, for massive IoT, frequent battery replacement leads to a prohibitively high human cost and environmental strain. In summary, massive access presents many challenging unsolved issues, which cannot be addressed with traditional approaches.

We note that the characteristics of massive access for cellular IoT are very different from those of the other two 5G use cases, namely enhanced mobile broadband (eMBB) and ultra-reliable low-latency communication (URLLC) \cite{5G5}. In particular, eMBB aims to provide high data rates for broadband applications such as virtual reality (VR) or argument reality (AR), while the objective of URLLC is to guarantee ultra-reliable low-latency communications for critical missions such as assisted/autonomous driving. Hence, for eMBB and URLLC, OMA schemes are preferred to achieve high spectral efficiency and link reliability. As pointed out above, 5G NB-IoT also employs OMA, but as a result, it cannot fully realize the goal of massive access \cite{5G2}. For example, NB-IoT can only accommodate fifty thousand devices per cell supporting a low data rate \cite{NB-IoT1}. For this reason, the ambitious goals of 5G NB-IoT have to be realized by B5G cellular IoT. Without doubt, the biggest challenge for B5G cellular IoT is the design of effective multiple access schemes that meet the corresponding performance requirements and services characteristics. In Table \ref{Tab2}, we compare the performance requirements of 5G NB-IoT and B5G cellular IoT. Compared to 5G NB-IoT, B5G cellular IoT imposes much more stringent requirements on power, connectivity, and coverage. Achieving these performance requirements using traditional multiple access techniques is very challenging. For example, it is challenging to realize wide coverage with low transmit power. Hence, new theoretical concepts, protocols, and techniques have to be developed for B5G cellular IoT to realize massive access.

The research on B5G cellular IoT has already begun in academia and industry. In \cite{Protocol4}, possible multiple access protocols for B5G cellular IoT were surveyed, with a focus on grant-free random access protocols based on approximate message passing (AMP) algorithms. Massive multiple-input multiple-output (MIMO) techniques for supporting cellular IoT were reviewed in \cite{MassiveMIMO00}, and corresponding research opportunities and challenges were identified. Moreover, as a promising approach for B5G cellular IoT, non-orthogonal multiple access (NOMA) was discussed in detail in \cite{NOMA0}. A common viewpoint of previous research is that B5G cellular IoT should further exploit degrees of freedom in the spatial, frequency, and user domains to facilitate significant performance improvements \cite{B5G1}-\cite{B5G3}. Generally speaking, previous survey papers have focused on one particular perspective of massive access, but do not provide a comprehensive overview of massive access in B5G cellular IoT which requires the consideration of many different aspects. To accelerate the development of massive access for the forthcoming B5G wireless networks, a comprehensive survey of the existing results, which can serve as building blocks for new research on next-generation cellular IoT, is necessary.

\begin{table}\centering
\caption{Comparison of 5G NB-IoT and B5G cellular IoT \cite{B5G4}.}
\label{Tab2}
\scriptsize
\begin{tabular}{|c|c|c|}\hline
  & 5G NB-IoT & B5G Cellular IoT \\
\hline
\hline Connectivity & \tabincell{l}{50 thousand per cell} & \tabincell{l}{10 million per km$^2$}\\
\hline Battery life & 10 years & 20 years \\
\hline Coverage & Ground  & Space-air-ground-sea \\
\hline Latency & 1 ms & 0.3 ms \\
\hline Reliability & $10^{-4}$ & $10^{-6}$ \\
\hline Positioning & 100 m & 1 m for outdoor and 10 cm for indoor  \\
\hline
\end{tabular}
\end{table}

\begin{figure*}[ht]
\centering
\includegraphics [width=0.8\textwidth] {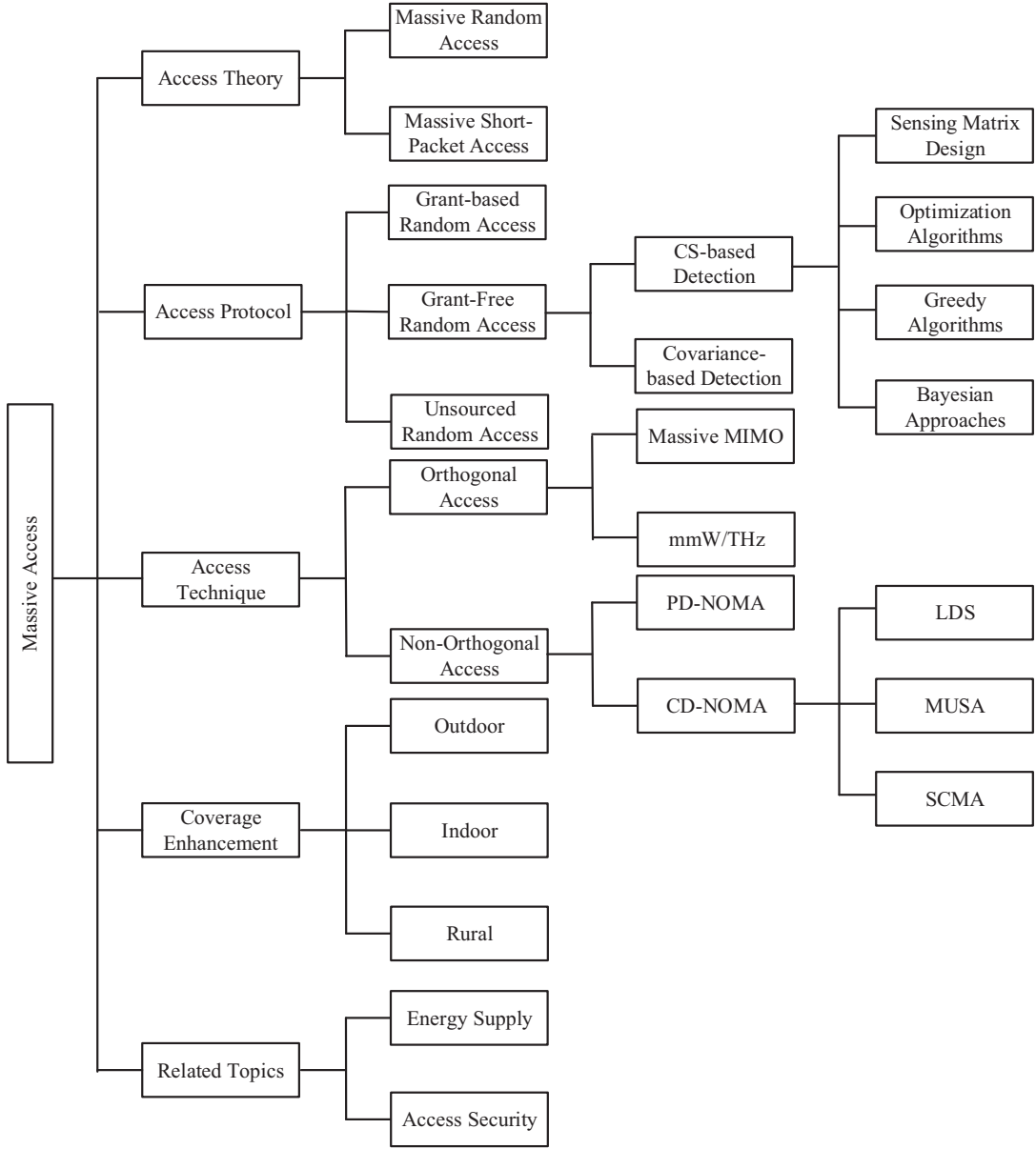}
\caption {Illustration of the aspects of massive access in B5G wireless networks considered in this survey paper. The labels of the rectangles correspond to the (sub)section titles.}
\label{Fig1}
\end{figure*}

The objective of this paper is to provide such a comprehensive overview of the latest results and progress on massive access in B5G wireless networks, c.f. Fig. \ref{Fig1}. The remainder of this paper is organized as follows. Section II introduces massive access in cellular IoT. Then, Section III investigates massive access from the perspective of information theory. Massive access protocols, massive access techniques, and massive coverage enhancement are discussed in Sections IV, V and VI, respectively. Moreover, energy supply for massive access and massive access security are considered in Section VII. Furthermore, future potential research directions for massive access are described in Section VIII. Finally, Section IX concludes the paper.

\section{Massive Access in Cellular IoT}
\begin{figure}[ht]
\centering
\includegraphics [width=0.46\textwidth] {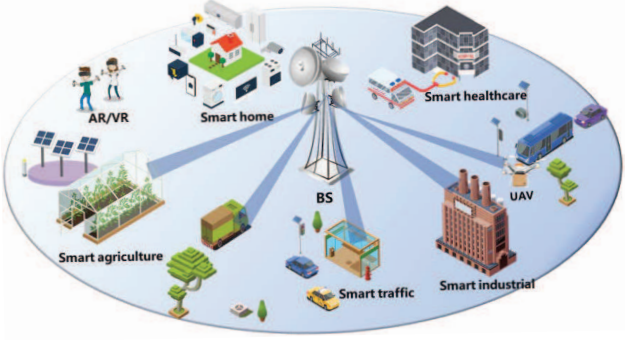}
\caption {Cellular IoT based on B5G wireless networks will be applied in various fields, e.g., industry, agriculture, traffic, and medicine.}
\label{Fig2}
\end{figure}

Wireless access refers to the last-mile connection between distributed end devices and a central station (e.g., a BTS). Due to the limited radio spectrum, multiple wireless devices have to share the same bandwidth employing multiple access techniques. In general, the performance of multiple access is determined by various factors such as channel conditions and device requirements. First, the wireless channel may experience fading, interference, and noise, which can significantly affect both access efficiency and reliability. Second, since multiple access schemes have to coordinate multiple devices, the quality-of-service (QoS) requirements of the devices, e.g., rate and latency, affect the selection of suitable access protocols and techniques. Hence, the design of multiple access is always a nontrivial issue. In B5G cellular IoT, c.f. Fig. \ref{Fig2}, the evolution from multiple access to massive access is driven by not only the envisioned massive number of IoT devices, but also the following critical service characteristics \cite{5G3}, \cite{5G4}:

\begin{itemize}

\item Sporadic traffic: IoT devices do not always have data to transmit creating bursty wireless traffic. In order to save energy, idle devices do not access the network. In general, a random number of devices access the network in each time slot.

\item Small payload: Most IoT applications infrequently generate small volumes of data having different sizes. In order to improve the resource utilization efficiency, short-packet transmission is preferable.

\item Low power: Ideally, the batteries of IoT devices should last for more than 20 years. Therefore, IoT devices have to employ an intelligent transmit power strategy to reduce the power consumption.

\item Ubiquitous distribution: In order to support various applications, IoT devices are distributed over a wide range, not only in urban areas, but also in rural areas. Hence, wide wireless coverage is needed.

\item Limited capability: Most IoT devices are wireless nodes with simple architecture, and limited/no energy storage. In other words, IoT devices cannot afford sophisticated signal processing operations.

\item Stringent latency constraint: Some IoT applications impose stringent latency requirements. Low-latency access schemes are needed to satisfy the latency requirements of such IoT applications.

\item Heterogenous QoS requirements: IoT devices across different application fields are very heterogeneous. For example, a small sensor for temperature sensing and a vehicle in a smart traffic system have very different QoS requirements, leading to different requirements for wireless access.

\end{itemize}

In general, massive access in B5G cellular IoT requires low power, massive connectivity, and broad coverage. Yet, the wireless channels for the last-mile connection between distributed end devices and the central station constitute a major bottleneck in meeting these performance requirements. First, the spectrum available in current wireless networks is limited. Second, the coherence time of wireless channels is also limited. For mobile applications such as smart traffic, the coherence time becomes very short \cite{Coherence0}. The length of the coherence time constrains the length of a data frame, which limits the performance of massive access. Moreover, for short coherence times, it is difficult to obtain full channel state information (CSI) for massive IoT and in some extreme scenarios with high mobility, CSI may not be available at all. In such cases, non-coherent transmission which does not require CSI may be adopted \cite{Coherence1}. However, non-coherent transmission suffers from a performance degradation compared to ideal coherent transmission \cite{Coherence2}. In summary, its unique characteristics and the properties of the underlying wireless channels lead to many challenging issues for realizing massive access. In the following sections, we introduce potential solutions from the perspectives of theories, techniques, and coverage enhancement.

\section{Massive Access Theories}
\begin{figure}[t]
\centering
\includegraphics [width=0.4\textwidth] {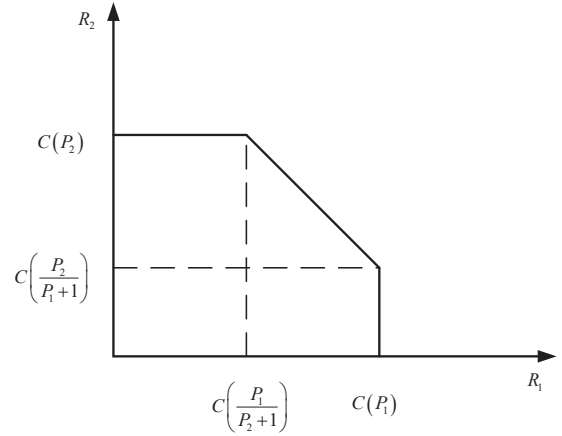}
\caption {The capacity region of the two-transmitter MAC.}
\label{Fig20}
\end{figure}
Information theory is the foundation of modern communications and can provide useful guidelines for the design of emerging wireless communication systems. To embrace the challenges introduced by massive access, we first revisit the capacity of the classical multiple access channel (MAC). For the conventional MAC, the channel capacity has been extensively studied \cite{IT3}-\cite{IT5}. It has been proven that the capacity region of a $K$-transmitter MAC with unit channel gain from each transmitter to the receiver can be characterized by \cite{IT6}
\begin{eqnarray}
\sum_{k \in \mathcal{S}} R_k < C\left(\sum_{k \in \mathcal{S}} P_k\right), \quad \forall \mathcal{S} \subseteq \{1, \cdots, K \},\label{eqn2}
\end{eqnarray}
where $R_k$ and $P_k$ are the $k$th transmitter's data rate and transmit power, respectively. The variance of the noise is normalized to 1 and $C(x)=\log_2(1+x)$ is the Shannon capacity formula. In Fig. \ref{Fig20}, we plot the capacity region of the two-transmitter MAC. The MAC capacity can be achieved by performing successive interference cancelation (SIC) at the receiver. For example, to achieve one corner point of the capacity region, the receiver first decodes the signal of one transmitter treating the signals of all other transmitters as noise and removes the decoded signal from the received signal. Then, the receiver decodes the next transmitter's signal treating the signals of the remaining transmitters as noise and removes the decoded signal again, until all signals are recovered. Note that for the MAC, the average per-transmitter channel capacity, i.e., $\frac{1}{K}\sum_{k=1}^{K}R_k$, asymptotically approaches zero as the number of transmitters $K$ tends to infinity. This is because co-channel interference becomes dominant when there is a massive number of transmitters. The conventional MAC capacity theory is only applicable for a fixed and finite number of transmitters. In fact, the above MAC capacity region is derived assuming infinitely long codes requiring a very large number of channel uses, which is incompatible with the typical IoT use cases \cite{IT6}. Considering the characteristics of B5G cellular IoT, an information theoretical study of massive access has to take into account the following requirements:
\begin{itemize}

\item Massive connectivity: There is a massive number of IoT devices with a density of more than $10$ million devices per km$^2$.

\item Random access: The sporadic traffic generated by typical IoT applications leads to random activity of the devices. Only active devices request access to the cellular network.

\item Short-packet transmission: The small payload of IoT data requires short-packet transmission to achieve high resource efficiency and low access latency.

\end{itemize}

Hence, realizing massive access in practical systems requires the development of new information theoretic design guidelines, which are quite different from the results available for the conventional MAC. Recently, the capacity of the massive access channel has been derived. In the following, we review these primary results.

\subsection{Massive Random Access}
The capacity of the massive access channel was first studied in \cite{IT7} based on the new notion of the many-access channel (MnAC). The MnAC model studies the scenario in which the number of transmitters increases unboundedly with the blocklength of the applied forward error correction (FEC) codes, both tending to infinity. Specifically, by applying random coding at the transmitters and Feinstein's threshold decoding at the receiver, as long as the number of transmitters $K$ grows sublinearly with the coding blocklength $M$, under a maximum power constraint $P$, each transmitter can send to the receiver a message of length
\begin{equation}
v(M)=\frac{M}{K}C(KP)\label{eqn3}
\end{equation}
bits with an arbitrarily small error probability if $M$ is sufficiently large. Note that $v(M)$ in (\ref{eqn3}) is a symmetric capacity since all transmitters achieve the same capacity. This result addresses the limitation of the conventional MAC capacity theory in analyzing the capacity of massive access for the case when the number of transmitters and the blocklength both go to infinity.

In \cite{IT7}, the symmetric capacity of the MnAC for the case of known active transmitter information is derived. Equivalently, this corresponds to the scenario where the transmitters are always active. However, as mentioned above, the transmitters in massive access are expected to be randomly active due to their sporadic traffic. For the case of random activity, practical decoding schemes adopted at the receiver have to involve two stages. The first stage identifies the set of active transmitters based on the superposition of their unique signatures (this corresponds to the active device detection problem in grant-free random access as will be discussed in Section IV). The second stage decodes the messages of the identified active transmitters. Intuitively, activity identification may lead to a loss in channel capacity. In \cite{IT8}, it is proven that for the MnAC with random activity, by using random coding at the transmitters and maximum-likelihood decoding at the receiver, if the number of transmitters $K$ grows as fast as linearly with the coding blocklength $M$, the symmetric capacity of a transmitter is given by
\begin{equation}
w(M)=\left(\frac{M}{\alpha K}C(\alpha KP)-\frac{H_2(\alpha)}{\alpha}\right)^+,\label{eqn4}
\end{equation}
where $(x)^+$ is defined as the maximum of $x$ and 0, $0\leq\alpha\leq1$ is the probability that a transmitter is active, and $H_2(\alpha)=-\alpha\ln(\alpha)-(1-\alpha)\ln(1-\alpha)$. Note that the term $\frac{H_2(\alpha)}{\alpha}$ is the difference between the MnAC capacity with and without activity information. Hence, the cost of activity identification is equal to the entropy of the activity probability \cite{IT9}.

The above papers considered the case where all nodes are single-antenna devices. For massive access in B5G wireless networks, both the BTS and the IoT devices may employ multiple antennas for performance enhancement \cite{MassiveMIMO1}, \cite{MassiveMIMO2}. Specifically, deploying multiple antennas at the BTS is generally affordable and has become standard in modern communication systems. Hence, it is necessary to characterize the capacity of the MnAC for the multiple-input multiple-output (MIMO) case. It was shown in \cite{IT10} that when the number of transmitters grows unbounded with the coding blocklength, the asymmetric ergodic message-length capacity of transmitter $k$ is given by
\begin{eqnarray}
u_k(M)&=&c_k\mathbb{E}_{\mathbf{H}}\left\{\log_2\det\left(\mathbf{I}_{N_R}+\sum_{t\in\mathcal{A}}\mathbf{H}_t\mathbf{Q}_t\mathbf{H}_t^{\dag}\right)\right\}\nonumber\\
&&-\mu_kKH_2(\alpha)\label{eqn5}
\end{eqnarray}
where $\mathbf{H}^{\dag}$ is the conjugate transpose of $\mathbf{H}$, $\mathbb{E}_{\mathbf{H}}\{x\}$ denotes expectation with respect to random variable $\mathbf{H}$, $\det (\cdot)$ returns the determinant of an input matrix, and $\mathcal{A}$ is the set of active transmitters. Here, $\mathbf{H}_t\in\mathbb{C}^{N_R\times N_T}$ and $\mathbf{Q}_t\in\mathbb{C}^{N_T\times N_T}$ are the channel matrix from the $t$th transmitter to the receiver and the covariance matrix of the codeword, respectively, $c_k=\underset{{M\rightarrow\infty}}{\lim}M\mu_k$, and $\mu_k=\frac{\log L_k}{\sum\limits_{t\in\mathcal{A}}\log L_t}$, where $N_R$ is the number of antennas at the receiver, $N_T$ is the number of antennas at each transmitter, and $L_k$ is the number of messages of the $k$th transmitter. The first term on the right hand side of \eqref{eqn5} is the individual capacity of the $k$th transmitter if activity information is available at the receiver. Hence, the individual capacity is proportional to the sum capacity with a scaling factor $c_k$, which depends on the number of messages. The second term on the right hand side of (\ref{eqn5}) is the cost of activity identification which is independent of the numbers of antennas at the transmitters and the receiver.

\begin{table*}\centering
\caption{Summary of Information Theoretical Results for Massive Access Systems.}
\label{Tab3}
\scriptsize
\begin{tabular}{|l|l|l|}\hline
 Reference & System model & Results\\
\hline\hline X. Chen \emph{et al.} \cite{IT7} &  \tabincell{l} {Massive access with known activity, long packet, single antenna} & \tabincell{l} {$v(M)=\frac{M}{K}C(KP)$} \\
\hline X. Chen \emph{et al.} \cite{IT8} & \tabincell{l} {Massive access with unknown activity, long packet, single antenna} & \tabincell{l} {$w(M)=\left(\frac{M}{\alpha K}C(\alpha KP)-\frac{H_2(\alpha)}{\alpha}\right)^+$} \\
\hline W. Fan \emph{et al.} \cite{IT10} & \tabincell{l} {Massive access with unknown activity, long packet, multiple antennas} & \tabincell{l} {$u_k(M)=c_k\mathbb{E}_{\mathbf{H}}\left\{\log_2\det\left(\mathbf{I}_{N_R}+\sum_{t\in\mathcal{A}}\mathbf{H}_t\mathbf{Q}_t\mathbf{H}_t^{\dag}\right)\right\}-\mu_kKH_2(\alpha)$} \\
\hline G. Durisi \emph{et al.} \cite{IT11} & \tabincell{l} {Multiple access with known activity, short packet, single antenna} & $R^{*}(M,\epsilon,P)\triangleq\sup\left\{\frac{\log_2 L}{M}:\exists(M,\epsilon,P) \mathrm{code} \right\}$ \\
\hline G. Ozcan \emph{et al.} \cite{IT14} & \tabincell{l} {Multiple access with known activity, short packet, single antenna} & \tabincell{l} {$R^{*}(M,\epsilon,P)\approx C(P)-\sqrt{\frac{V}{M}}\frac{Q^{-1}(\epsilon)}{\ln2}$}\\
\hline
\end{tabular}
\end{table*}

\subsection{Massive Short-Packet Access}
The results in \cite{IT7}-\cite{IT9} and \cite{IT10} on massive random access are based on the common assumption that the coding blocklength or the number of channel uses increases in the same order as the number of transmitters. In this case, the packet error rate (PER) can approach zero as the coding blocklength tends to infinity. However, in practical systems, the coding blocklength is finite. Especially, for cellular IoT, short packets are preferred due to their lower latency for bursty data communications \cite{Shortpacket2}, \cite{Shortpacket3}. In the context of short-packet transmission, it is difficult to guarantee error-free reception for a short activity period. Thus, for a given packet length, short-packet transmission should achieve a balance between spectral efficiency and decoding error probability. Formally, the capacity of short-packet transmission, $R^{*}(M,\epsilon,P)$, can be defined as the largest rate $(\log_2 L)/M$ for which there exists an $(M,\epsilon,P)$ code, namely \cite{IT11}
\begin{equation}
R^{*}(M,\epsilon,P)\triangleq\sup\left\{\frac{\log_2L}{M}:\exists(M,\epsilon,P) \mathrm{code} \right\},\label{eqn6}
\end{equation}
where $\epsilon>0$ is the PER and $L$ is the number of messages. Via asymptotic analysis, it can be shown that (\ref{eqn6}) generalizes the well-known existing capacity results. For instance, as $M\rightarrow\infty$, it is equivalent to Shannon's capacity \cite{IT12}. Moreover, when $P$ tends to infinity, it is possible to obtain the diversity-multiplexing tradeoff proposed by Zheng and Tse \cite{IT13}. However, it is challenging to derive the exact capacity for massive short-packet access from (\ref{eqn6}) in closed form.

For the ease of analysis, a tight approximation for $R^{*}(M,\epsilon,P)$ was derived in \cite{IT14} as follows
\begin{eqnarray}
R^{*}(M,\epsilon,P)\approx C(P)-\sqrt{\frac{V}{M}}\frac{Q^{-1}(\epsilon)}{\ln2},\label{eqn7}
\end{eqnarray}
where $Q^{-1}(x)$ is the inverse Gaussian Q function, $Q(x)=\int_x^{\infty}\frac{1}{\sqrt{2\pi}}\exp\left(-\frac{t^2}{2}\right)dt$, and $V$ is the channel dispersion. Intuitively, as $M$ tends to infinity, (\ref{eqn7}) reduces to the Shannon capacity formula. Based on (\ref{eqn7}), one can evaluate the performance of massive short-packet access. Specifically, by substituting the signal-to-interference-plus-noise ratio (SINR) for massive short-packet access into (\ref{eqn7}), the achievable rate for each transmitter can be evaluated.

Generally speaking, the results available for the capacity of massive access in practical wireless networks are still very limited. Most existing theoretical works only consider Gaussian channels. If the channels suffer from fading and need to be estimated, the capacities in (\ref{eqn3}) and (\ref{eqn4}) for massive random access will be quite different. Furthermore, for massive short-packet access over fading channels, the capacity of short-packet transmission essentially reduces to the outage capacity \cite{IT11}. A summary of existing results on massive access information theory is given in Table \ref{Tab3}.

\section{Massive Access Protocols}
\begin{figure}[t]
\centering
\includegraphics [width=0.45\textwidth] {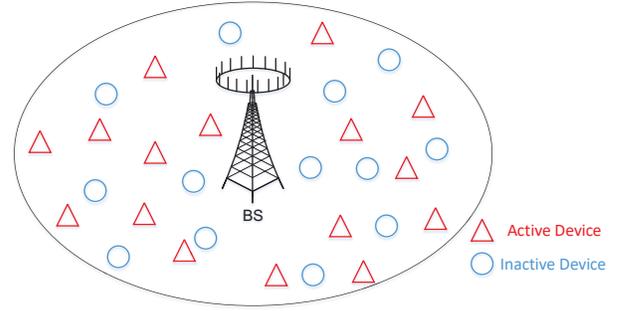}
\caption {Illustration of sporadic traffic of IoT applications. In general, during an arbitrary time slot, only a fraction of the devices has data to transmit, namely the active devices.}
\label{Fig3}
\end{figure}

Due to the sporadic traffic of IoT applications, only a fraction of the devices, namely the active devices, have data to transmit at a given time, as shown in Fig. \ref{Fig3}. Access protocols are used to coordinate the access requests of the active IoT devices \cite{Protocol1}. Specifically, each active device contacts the BTS to access the network. Then, the BTS identifies the active devices by some means. Hence, an access protocol is needed to coordinate the data exchange between the BTS and the IoT devices. In general, the activity of the IoT devices is random. Consequently, random access protocols are commonly used in cellular IoT \cite{Protocol2}, \cite{Protocol3}. Two types of random access protocols are commonly distinguished, namely grant-based and grant-free random access protocols \cite{Protocol4}. Moreover, a new random access protocol called unsourced massive random access has been proposed recently \cite{Unsourced1}. We discuss these three massive random access protocols in the following.

\subsection{Grant-Based Random Access}
\begin{figure}[ht]
\centering
\includegraphics [width=0.35\textwidth] {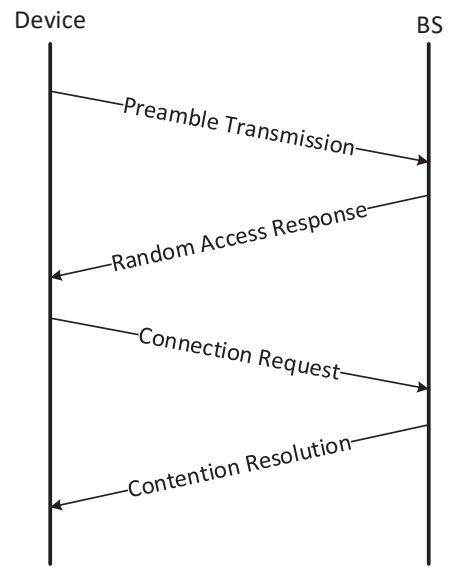}
\caption {Grant-based random access protocol.}
\label{Fig4}
\end{figure}

Grant-based random access is adopted in the current 5G NB-IoT \cite{Protocol4}. As the name implies, for a grant-based random access protocol, an active device needs to obtain permission from the BTS to access the network. As shown in Fig. \ref{Fig4}, the grant procedure of a typical grant-based random access protocol, such as ALOHA, includes four transmissions between the IoT device and the BTS as follows \cite{Protocol5}:
\begin{enumerate} [(1)]
\item Each active device randomly selects a preamble (also referred to as a signature) from a pool of orthogonal preamble sequences and uses the selected preamble to inform the BTS that it has data to transmit.

\item The BTS responds to each active device authorizing it to send a connection request in the next stage.

\item The active devices send connection requests for resource allocation for data transmission.

\item If a preamble is picked by only one active device, the BTS will grant the corresponding request and send a contention-resolution message to inform the active device about the allocated resources. Otherwise, the access request is not granted.

\end{enumerate}

\begin{table*}\centering
\caption{Preamble sequences for massive device detection.}
\label{Tab4}
\scriptsize
\begin{tabular}{|l|l|l|}\hline
 Reference & Preamble Sequence & Advantage\\
\hline\hline J. Ding \emph{et al.} \cite{Protocol11} &  ZC sequences & Good auto- and cross-correlation properties \\
\hline L. Liu \emph{et al.} \cite{Protocol12} & Gaussian sequences & Easy to generate and convenient for performance analysis \\
\hline J. Wang \emph{et al.} \cite{Protocol14} & RM sequences & Reduced storage space requirements \\
\hline S. Li \emph{et al.} \cite{Protocol140} & Deep auto-encoded sequences & Adaptive to sparse patterns even without analytical models \\
\hline
\end{tabular}
\end{table*}

The main advantage of the grant-based random access protocol is the simple processing at the BTS. However, in the context of massive access, the grant-based random access protocol has the following shortcomings. First, the number of orthogonal preamble sequences is finite due to the short coherence time. If there exist a massive number of IoT devices, the probability that a preamble is selected by more than one device is high. In other words, the devices suffer from a high probability of access failure due to collision. As a consequence, the average access latency may become too high to be tolerable. Second, the grant-based random access protocol requires four transmissions, resulting in a high signaling overhead. Since the channel capacity is limited, the required signaling overhead might be prohibitively large for massive access.

\subsection{Grant-Free Random Access}
To overcome the problems of grant-based random access, various grant-free random access protocols have been proposed which allow the active devices to access the wireless network without a grant \cite{Protocol6}. To be specific, active devices first send their unique preambles to the BTS and then transmit the data signals directly \cite{Protocol7}. Therefore, both access latency and signaling overhead are significantly reduced. The key idea of grant-free random access is to detect the active devices based on the received preambles at the BTS \cite{Protocol8}. For massive access, due to the massive number of devices and the use of short packets, the preamble sequences are not orthogonal. As a result, the received preamble signals suffer from severe co-channel interference. Hence, the BTS has to adopt sophisticated activity detection algorithms. In other words, grant-free random access reduces the access delay and the signaling overhead at the expense of a high computational complexity at the BTS. In general, grant-free random access for massive access requires massive device detection at the receiver. This can be done using a compressed sensing (CS)-based sparse signal recovery framework or a covariance-based approach for massive device detection as discussed below.

\subsubsection{CS Formulation}
Due to the sporadic traffic generated by IoT applications, the received preamble signal is typically sparse when only the active devices send their preambles. It is well known that the resulting sparse signal recovery problem from noisy measurements can be tackled with CS methods \cite{Protocol9}, \cite{Protocol10}. The CS problem for massive device detection can be formulated as
\begin{eqnarray}
&\underset{{\mathbf{X}}}{\mathrm{minimize}}\quad\|\mathbf{X}\|_{0}\nonumber\\
&\quad\quad\quad\quad\quad\mathrm{s.t.}\quad\|\mathbf{Y}-\mathbf{AX}\|_{\mathrm{F}}\leq\delta,\label{eqn8}
\end{eqnarray}
where $\|\cdot\|_0$ is the zero-norm defined as the number of nonzero elements of the argument and $\|\cdot\|_{\mathrm{F}}$ is the Frobenius norm. In the above CS problem, $\mathbf{Y}$ is the space-time received preamble signal and $\delta$ is a predetermined error tolerance constant which depends on the noise power. Moreover, $\mathbf{A}=[\mathbf{a}_1,\cdots,\mathbf{a}_K]$ is the sensing matrix with $\mathbf{a}_k$ being the $k$th device's preamble sequence, and $\mathbf{X}=[\alpha_1\mathbf{h}_1,\cdots,\alpha_K\mathbf{h}_K]^T$ is the device state matrix with $\alpha_i$ and $\mathbf{h}_i$ being the activity indicator and the channel response of the $i$th device, respectively. Here, $\alpha_i=1$ if the $i$th device is active, otherwise $\alpha_i=0$. Hence, multiple rows of $\mathbf{X}$ are zero and the aim of the CS problem (\ref{eqn8}) is to determine the nonzero rows of $\mathbf{X}$ from the noisy measurements $\mathbf{Y}$, i.e., activity detection. The CS problem (\ref{eqn8}) is generally nonconvex and thus it is difficult to obtain the globally optimal solution directly and efficiently. Therefore, massive device detection algorithms are usually designed based on a relaxed CS problem. In what follows, we discuss different aspects of the design of CS-based massive device detection algorithms.

\emph{[a] Sensing Matrix Design}: The sensing matrix $\mathbf{A}$ has a significant impact on the design of massive device detection algorithms and hence determines the performance of grant-free random access, namely the detection probability. Since the preamble sequences are nonorthogonal, the design of the sensing matrix is not a trivial task. In 4G LTE systems, Zadoff-Chu (ZC) sequences are adopted as preamble signals because of their good auto- and cross-correlation properties. ZC sequences were used as preamble sequences for grant-free random access systems in \cite{Protocol11} and their performance was compared to that of orthogonal sequences. Recently, independent and identically distributed (i.i.d.) Gaussian sequences have been considered as preamble sequences to study grant-free random access. This is because Gaussian sequences can be easily generated and are convenient for performance analysis. For instance, the detection probability for Gaussian distributed preamble sequences was derived in \cite{Protocol10} and \cite{Protocol12}. It was shown that the detection probability improved as the number of BTS antennas was increased. Theoretically, the active devices can be detected perfectly in the asymptotic limit as the number of BTS antennas goes to infinity. Furthermore, the impact of the length of Gaussian distributed preamble sequences on the detection probability was analyzed in \cite{Protocol13}. Then, the durations of the preamble and data sequences in a frame were optimized to maximize the system spectral efficiency. Since each device is assigned a unique preamble sequence, the BTS has to allocate a large amount of storage capacity to store the preamble sequences in the massive access case. In order to reduce the required storage space, Reed-Muller (RM) sequences can be applied as preamble sequences. The authors in \cite{Protocol14} exploited the nested structure of RM sequences and their sub-sequences to design a low-complexity activity detection algorithm. Moreover, a data-driven deep learning method was applied to generate preamble sequences, which can adapt to wireless channels with arbitrary distribution. In \cite{Protocol140}, a deep auto-encoder was utilized to jointly design preamble sequences and the corresponding sparse signal recovery algorithm, which can effectively exploit sparsity patterns even without analytical models. Several potential preamble sequences for massive device detection are compared in Table \ref{Tab4}.

\emph{[b] CS Algorithms}: Since the zero-norm in the objective function of the CS optimization problem (\ref{eqn8}) is nonconvex, it is impossible to design massive device detection algorithms by solving the original problem optimally with polynomial time computational complexity \cite{CS1}. In the literature, optimization algorithms, greedy algorithms, and Bayesian approaches have been utilized to obtain effective suboptimal solutions for the above CS optimization problem in the context of massive device detection. These algorithms are explained in the following.

\emph{[b.1] Optimization Algorithms}: In order to obtain a feasible solution of the CS problem (\ref{eqn8}), it is necessary to approximate the objective function. In \cite{OA1}, the zero-norm $\|\mathbf{X}\|_{0}$ was replaced by the sum of all entries of $\mathbf{X}$. Thus, the original problem was transformed to a linear programming problem which can be solved optimally with low complexity. The solution of the linear programming problem can be proved to be identical to that of the original problem only when $\mathbf{X}$ is sufficiently sparse and there is no noise. On the other hand, since the $l_1$-norm is convex, many papers base the algorithm design on $l_1$-regularization problems. For example, the authors in \cite{OA2} proved that if $\mathbf{X}$ is sufficiently sparse, $l_1$-regularization problems can accurately recover $\mathbf{X}$ even for noisy measurements. Furthermore, the authors in \cite{OA3} transformed the original CS problem to an $l_1$-regularization least-squares problem and proposed a customized interior-point method for solving the problem.

For massive access in B5G wireless networks, the number of IoT devices and the number of BTS antennas are expected to be very large, resulting in a high dimensional device state matrix $\mathbf{X}$. Hence, even with $l_1$-regularization, the computational complexity may still be prohibitive. In fact, due to the spatial correlation of the BTS antennas, $\mathbf{X}$ may not only be sparse, but also low-rank. The simultaneous sparsity and low-rank property can be exploited to further decrease the computational complexity of activity detection. In \cite{OA4}, a rank or nuclear norm constraint was inserted into the $l_1$-regularization problem. Theoretical analysis revealed that such a nuclear norm constrained problem can achieve near-optimality based on a small number of measurements. To further decrease the computational complexity and the required length of the preamble sequences, a rank-aware $l_1$-regularization least-squares problem was formulated by estimating the rank of $\mathbf{X}$ in advance \cite{OA5}. For a given rank, the $l_1$-regularization least-squares problem can be transformed into a low-dimensional problem. Yet, the rank-aware problem is usually nonconvex. To tackle this challenge, a Riemannian optimization-based algorithm was proposed in \cite{OA5} to obtain a suboptimal solution.

Moreover, $l_p$-norm minimization with $0<p<1$ can be used to develop optimization algorithms for massive device detection \cite{CS1}. Since the $l_p$-norm is a better approximation for the $l_0$-norm than the $l_1$-norm, $l_p$-norm minimization may achieve more accurate activity detection. However, due to the nonconvexity of the $l_p$-norm, $l_p$-norm minimization usually leads to a high computational complexity.

\emph{[b.2] Greedy Algorithms}:
To avoid having to solve nonconvex CS problems via non-polynomial time algorithms, greedy algorithms can be applied to massive device detection in an effort to reduce the computational complexity at the expense of a degradation in performance. Greedy algorithms are iterative approaches that take local optimal decisions in each step to eventually obtain an effective suboptimal solution. One of the most widely used greedy algorithms in device detection is the group orthogonal matching pursuit (GOMP) algorithm \cite{gomp}. Since the device state matrix $\mathbf{X}$ in general contains only a few nonzero rows, in the absence of noise, the measurements are in a space supported by a sub-matrix of $\mathbf{A}$ determined by the positions of these nonzero rows. The GOMP iteratively builds the support of $\mathbf{X}$ and enhances the device detection performance by accumulating the correlation between the residual and the sensing matrix. Although the GOMP detection algorithm based on correlation is simple, its performance is heavily affected by noise. Another type of greedy algorithm, namely the Hierarchical Hard Thresholding Pursuit (HiHTP) algorithm \cite{HIHTP,HIHTP1}, makes use of not only the block structure of the device state matrix, but also the intra-block sparse structure caused by the channel taps. The sporadic device activity and the sparse channel profiles give rise to a hierarchically sparse structured vector containing all estimated channel coefficients. Motivated by this observation, a prediction of the support of the device state matrix can be inferred by applying a thresholding operation based on the hierarchically sparse structure. Then, the best $l_2$-norm approximation to the received signal compatible with this support is calculated. The HiHTP algorithm can efficiently reconstruct hierarchically sparse signals from only a small number of linear measurements.

One important property of greedy detection algorithms is their simplicity of implementation. However, a drawback is their inherent error propagation, since previous choices for device activity are not re-evaluated. Moreover, the detection performance of these algorithms is seriously affected by the noise level.

\begin{table*}\centering
\caption{Massive Device Detection Algorithms.}
\label{Tab5}
\scriptsize
\begin{tabular}{|l|l|l|}\hline
 Reference & Algorithm Type & Description\\
\hline\hline M. Golbabaee \emph{et al.} \cite{OA4} &  \tabincell{l} {Optimization algorithm} & \tabincell{l} {Rank-constrained $l_1$-regularization optimization algorithm} \\
\hline X. Shao \emph{et al.} \cite{OA5} & \tabincell{l} {Optimization algorithm} & \tabincell{l} {Rank-aware Riemannian optimization algorithm} \\
\hline C. Bockelmann \emph{et al.} \cite{gomp} & \tabincell{l} {Greedy algorithm} & \tabincell{l} {Group orthogonal matching pursuit algorithm} \\
\hline I. Roth \emph{et al.} \cite{HIHTP} & \tabincell{l} {Greedy algorithm} & Hierarchical hard thresholding pursuit algorithm \\
\hline Z. Chen \emph{et al.} \cite{Protocol10} & \tabincell{l} {Bayesian algorithm} & \tabincell{l} {Approximate message passing algorithm}\\
\hline M. Ke \emph{et al.} \cite{GMMV-AMP} & \tabincell{l} {Bayesian algorithm} & \tabincell{l} {Generalized multiple measurement vector approximate message passing algorithm}\\
\hline
\end{tabular}
\end{table*}

\emph{[b.3] Bayesian Approaches}:
To improve the detection performance of massive random access, Bayesian CS-based detection algorithms have been developed, e.g. \cite{CRAN}-\cite{GMMV-AMP}. This kind of detection algorithm first assigns a prior probability distribution which promotes sparsity to the unknown device state matrix, and then infers the posterior distribution of the unknown signal from the received signal at the BTS. By exploiting the prior channel information regarding the path loss and the chunk sparsity structure, the authors in \cite{CRAN} proposed a Bayesian CS-based algorithm to efficiently detect device activity in an uplink cloud radio access network. However, the Bayesian formulation in \cite{CRAN} was developed based on the assumption of infinite-capacity fronthaul links, which is not practical. Thus, taking into account the impact of fronthaul capacity limitations, the authors in \cite{CRAN1} employed a hybrid generalized approximate message passing (GAMP) method, which was based on a quadratic approximation of the sum-product message passing scheme and accommodated both nonlinear measurements and group sparsity to enhance the device detection performance.

To further exploit statistical channel knowledge, the authors in \cite{Protocol10} adopted a Bayesian approach where the sparsity was modeled via the prior distribution of the channel to facilitate the development of an improved version of the approximate message passing (AMP) algorithm. The authors in \cite{Protocol12} further demonstrated that in an asymptotic regime where the number of users, the pilot length and the number of BTS antennas all go to infinity in a particular manner, both the miss detection and the false alarm probabilities of the AMP algorithm for activity detection can asymptotically approach zero. The exact knowledge of the prior distribution of the channels and the noise variance may be difficult to obtain in practice due to the sporadic traffic and the spatial correlation of the channels. Furthermore, the above works considered massive device detection in narrowband scenarios. In fact, B5G wireless networks might employ broadband systems, e.g., millimeter wave (mmW) or even terahertz (THz) systems \cite{Wideband}. Towards this end, the authors in \cite{GMMV-AMP} proposed a generalized multiple measurement vector approximate message passing (GMMV-AMP) algorithm to adaptively detect the active devices by exploiting the virtual angular domain sparsity of the channels in an orthogonal frequency division multiplexing (OFDM) broadband system. Furthermore, the expectation maximization (EM) algorithm was utilized to learn the unknown hyper-parameters of the channel and noise distributions. Several massive device detection algorithms are compared in Table \ref{Tab5}.


\emph{[c] Joint Device Detection and Channel Estimation}: To realize effective massive access, the BTS requires accurate CSI for decoding the uplink signals and performing precoding of the downlink signals after activity detection. In general, CSI is acquired through channel estimation at the BTS based on pilot sequences sent by the devices. Since the preamble sequences for activity detection can be also exploited as pilot sequences for channel estimation, activity detection and channel estimation can be jointly performed based on the same sequences.

Recently, several joint activity detection and channel estimation (JADCE) algorithms for massive connectivity in cellular IoT networks have been reported. Since JADCE is still a CS problem, the authors in \cite{JADCE} proposed a Lasso-based $l_{2,1}$-regularization penalty function to exploit the inherent sparsity existing in both the device activity and the remote radio heads with which the active devices are associated. Then, an alternating direction method of multipliers (ADMM) algorithm was applied to handle the resulting large-scale convex JADCE problem. In fact, Bayesian algorithms can be also adopted to handle the JADCE problem. In \cite{MassiveMIMO7}, by exploiting statistical information about the wireless channels, a JADCE algorithm was designed for massive MIMO systems to jointly detect device activity and estimate the CSI. Furthermore, an expectation propagation (EP)-based JADCE algorithm was proposed in \cite{EP} for massive access. This algorithm approximated the computationally intractable probability distribution of the sparse channel vector by an easily tractable distribution, which can substantially enhance JADCE performance.

A major problem of the above JADCE algorithms is that in practical scenarios with short pilot sequences, their performance is severely degraded. To tackle this problem, the authors in \cite{control} proposed a transmission control scheme for grant-free random access protocols. Specifically, based on a predetermined transmission control function, each active device decides to transmit a packet in the current time slot or to postpone the transmission. At the BTS, a modified AMP algorithm was adopted to improve JADCE performance. Transmission control was motivated by the fact that decreasing the number of active devices can significantly improve JADCE performance for a given pilot length.

\emph{[d] Joint Device and Data Detection}: To reduce access latency and signaling overhead, blind detection has become a promising approach to jointly detecting devices and data for massive access scenarios without prior knowledge of the CSI, especially for low-latency communications. For instance, the authors in \cite{Protocol9} proposed a non-coherent transmission scheme that does not need CSI at the BTS and developed a modified AMP algorithm to exploit the structured sparsity caused by the scheme. For this algorithm, explicit channel estimation is not required because of the non-coherent transmission, and the data signal is embedded into the pilot sequences. Motivated by the observation that if the active devices transmit symbols that are either $-1$ or $1$ and the inactive devices are modelled as transmitting all-zero symbols, the transmit symbol alphabet is ternary, the authors in \cite{ternary} proposed an information-enhanced adaptive matching pursuit algorithm for joint device and data detection. Moreover, the authors in \cite{MAP} proposed a maximum a posteriori probability (MAP)-based device and data detection algorithm, which comprises a MAP-based active user detector (MAP-AUD) and a MAP-based data detector (MAP-DD). Extrinsic information is exchanged between the MAP-AUD and the MAP-DD. In particular, joint detection of the active devices and the data symbols is performed first, then the estimated data symbol is refined and used as a priori information for the detection of the active devices.

The above algorithms carry out joint device and data detection within one time slot. In other words, they do not exploit the temporal correlation across time slots. To this end, a dynamic CS-based device and data detector using orthogonal matching pursuit (OMP) across time slots was proposed in \cite{AD1}. Moreover, an a priori information aided adaptive subspace pursuit (PIA-ASP) algorithm was proposed in \cite{AD2} to detect active devices and data symbols. In the PIA-ASP algorithm, a parameter evaluating the quality of the prior-information support set was introduced, so as to exploit the intrinsic temporal correlation of the active device support sets across several continuous time slots.

\subsubsection{Covariance Formulation}
If we are only interested in detecting the device activities and not interested in estimating the channel and if the BTS is equipped with a large number of antennas, it is possible to formulate the massive device detection as a maximum likelihood estimation problem based on the covariance matrix of the received signal at the BTS, and then employ a coordinate descent method to obtain a suboptimal solution \cite{Covariance1}, \cite{Unsourced3}. The key advantage of this covariance-based approach is that it is able to detect a much larger number of active devices. In fact, the number of active devices can scale quadratically with the length of the pilot sequences, thereby alleviating a key bottleneck in massive access. This scaling law was established under a so-called non-negative least square (NNLS) formulation in \cite{Unsourced3}, and can also be analyzed via the Fisher information matrix of the maximum likelihood problem \cite{Covariance2}. We note that the covariance-based approach can also be used for joint activity and data detection. Specifically, each device is assigned not only one sequence but a unique sequence set. The transmitted sequence corresponds to the transmitted data. Hence, by detecting the received sequence, the activity information and the data information can be obtained simultaneously.

Generally speaking, massive device detection for grant-free random access is still an open problem, which involves two challenging unsolved issues. First, the existing algorithms entail a high computational complexity for recovering the device state matrix due to its large-dimensional structure. Second, the required length of the preamble sequences may be too long for short-packet transmission in B5G wireless networks.

\subsection{Unsourced Random Access}
Recently, a new massive random access paradigm, referred to as unsourced massive random access, was proposed in \cite{Unsourced1}. Unlike grant-based and grant-free random access protocols that assign each device a unique preamble sequence, unsourced massive random access utilizes one codebook (a set of sequences) for all devices. The devices include their identity (ID) in the information message itself, and the BTS decodes the list of active device messages up to permutations. It has been shown that unsourced massive random access can significantly decrease the minimum energy per bit required for reliable communication. The authors in \cite{Unsourced2} extended unsourced massive random access to the case where the BTS had a very large number of antennas and no CSI. Specifically, the minimum energy required for reliable communication can be made arbitrarily small as the number of BTS antennas grows sufficiently large. However, there are many challenging unsolved problems in unsourced massive random access, such as efficient codebook design and activity detection algorithms. Some recent progress on codebook design for massive access has been reported in \cite{Unsourced4}.

\section{Massive Access Techniques}
Access techniques organize the data exchange between the active devices and the BTS. In previous generations of cellular networks, OMA techniques, such as TDMA, FDMA, and OFDMA, have been adopted. For 5G NB-IoT, SC-FDMA is employed for the uplink and OFDMA for the downlink. In particular, OMA techniques allocate each time-frequency resource block to a unique device which leads to a simple transceiver structure. However, due to the limited radio spectrum available for cellular communications, it is difficult to support massive access with the conventional OMA techniques. To tackle this challenge, there are two possible directions for massive access in B5G wireless networks. On the one hand, new wireless resources, e.g., new spectrum, can be utilized to admit more devices. On the other hand, resource utilization efficiency can be further improved to support massive access. In the following, we describe two different massive access techniques which are along the above-mentioned directions.

\subsection{Massive Orthogonal Access}
Conventional OMA techniques over limited radio spectrum cannot satisfy the stringent QoS requirements of massive access, and hence B5G wireless networks have to adopt new massive orthogonal access techniques by exploiting extra degrees of freedom. Due to the strict latency constraints, time-domain resources are scarce and cannot be used for massive access. Instead, exploiting additional space- and frequency-domain resources is more attractive.

\subsubsection{Massive MIMO}
Multiple-antenna techniques have been adopted in 4G long-term evolution (LTE) networks to increase transmission rate and to enhance link reliability by exploiting extra spatial degrees of freedom. However, the BTSs of 4G LTE can be equipped only with up to eight antennas. Thus, the spatial degrees of freedom offered by LTE multiple-antenna BTSs are limited and far from enough to facilitate massive access. To significantly increase the available spatial degrees of freedom, the BTSs of B5G wireless networks will deploy a large-scale antenna array with 64 or more antennas, realizing massive MIMO. Therefore, a large number of devices can access the network simultaneously and the BTS can separate them in the spatial domain, e.g. space division multiple access (SDMA) \cite{MassiveMIMO3}, \cite{MassiveMIMO4}.

The pioneering work on massive MIMO in \cite{MassiveMIMO5} showed that co-channel interference vanishes asymptotically even with simple linear precoders and combiners as the number of BTS antennas tends to infinity due to channel hardening. Moreover, it was shown that both the spectral and energy efficiencies can be improved significantly by using massive MIMO \cite{MassiveMIMO6}, \cite{MassiveMIMO61}. In the case of massive access, massive MIMO does not only increase the accuracy of active device detection but also improves the transmission performance \cite{MassiveMIMO7}. However, there are two critical issues for implementing massive access in massive MIMO systems.

The first issue concerns the CSI acquisition at the BTS. The accuracy of the CSI at the BTS determines the performance of massive access based on massive MIMO \cite{MassiveMIMO8}. Since the BTS is at the transmitter side for the downlink, it is impossible to obtain CSI directly. In traditional multiple-antenna systems, there are two CSI acquisition methods. In frequency division duplex (FDD) systems, the devices first obtain the CSI by channel estimation and then feed back the quantized CSI to the BTS \cite{MassiveMIMO9}. For massive access based on massive MIMO, the required number of feedback bits at each device is large due to the high dimensional channel vector. Thus, the total amount of feedback required for a large number of devices can be prohibitive. In other words, conventional quantized feedback methods are not applicable for massive access based on massive MIMO. However, if the channel is sparse, several effective methods, e.g. CS \cite{MassiveMIMO10}, \cite{MassiveMIMO11}, can reduce the amount of feedback. In \cite{MassiveMIMO12}, the received pilot signal of each device was conveyed to the BTS and the sparse CSI of all devices was jointly recovered by using a $l_1$-regularization-based CS method. Moreover, deep learning was employed to compress the sparse CSI in \cite{MassiveMIMO13}, such that the amount of feedback became affordable. The requirement of sparse CSI limits the applicability of the above methods to FDD systems. Hence, massive MIMO is usually envisioned to operate in the time division duplex (TDD) mode. In TDD systems, the devices send pilot sequences to the BTS in the uplink and the BTS obtains the downlink CSI by estimating the uplink channels exploiting channel reciprocity \cite{MassiveMIMO14}. An obstacle to realizing CSI acquisition in TDD systems is the so-called pilot contamination problem \cite{MassiveMIMO15}. Specifically, due to the limited pilot sequence length available for serving a massive number of devices, the same pilot sequences have to be reused in different devices. Consequently, the CSI estimation accuracy is reduced due to co-channel interference. Because the pilot sequences in massive access cannot be completely orthogonal, it is necessary to improve the CSI accuracy. To this end, in \cite{MassiveMIMO16}, a pilot transmit power control scheme was proposed, so as to improve the overall performance. Since the CSI accuracy depends on the pilot transmit energy, for a given transmit power, the pilot sequence length should be optimized as in \cite{Protocol13}.

The second issue concerns the energy consumption and the associated cost. The use of massive MIMO for massive access requires a large number of radio-frequency (RF) chains and the associated analog-to-digital converter (ADC) modules. If each antenna is equipped with a dedicated RF chain, the number of RF chains can be very large resulting in high energy consumption \cite{ng2012energy2}. To decrease the number of RF chains but retain the benefits of massive MIMO, hybrid precoding techniques are needed to allow multiple antennas to share the same RF chains \cite{MassiveMIMO17}--\nocite{MassiveMIMO18}\cite{JR:Zhiqiang_hybrid}. Compared to conventional digital precoding schemes, hybrid precoding incurs high design complexity but low implementation cost. In \cite{MassiveMIMO19}, a penalty dual decomposition-based hybrid precoding design method was proposed, which is guaranteed to converge to a Karush-Kuhn-Tucker (KKT) solution of the hybrid precoding problem under some mild assumptions. To decrease the design complexity, a two-timescale hybrid precoding method was presented in \cite{MassiveMIMO20}, which constructed the analog beamforming based on slowly time-varying statistical CSI. Moreover, the high cost of ADC is a vital issue for massive MIMO. Since the ADC cost is mainly determined by the resolution of the quantization, a low-resolution ADC is preferred for massive access based on massive MIMO at the cost of a performance loss \cite{MassiveMIMO21}, \cite{MassiveMIMO22}. In \cite{MassiveMIMO23}, the impact of low-resolution ADC on the performance of massive access was analyzed, and a time allocation scheme for channel estimation and data transmission was proposed to alleviate the impact of low-resolution ADC.

\subsubsection{Millimeter-Wave/Terahertz}
According to the Shannon capacity theorem, increasing the bandwidth is a simple but effective way for improving the capacity of wireless communications. Current cellular networks operate in sub-6 GHz bands which provide limited bandwidth and are overcrowded. On the contrary, high frequency bands have large vacant spectra. Hence, for B5G wireless networks, the use of mmW and even THz bands is attractive in order to realize massive access in the frequency domain \cite{mmWave1}-\cite{mmWave3}. A critical issue for mmW/THz communications is the severe propagation loss, resulting in a short transmission distance. To address this problem, mmW/THz is usually combined with massive MIMO or even ultra-massive MIMO employing more than one thousand antennas \cite{mmWave31}. Thus, CSI acquisition and precoding become more complicated for mmW/THz communications. Fortunately, mmW/THz channels have two important characteristics. Firstly, the mmW/THz channel is very sparse, such that CS and Bayesian methods can be used to acquire the CSI required for the design of the precoding matrix. In \cite{mmWave4}, to obtain CSI, the $l_{1,2}$-regularization-based CS method was applied, which can avoid quantization errors and provide super-resolution performance. By modeling the channel coefficients as Laplacian distributed random variables, a GAMP algorithm was used to find the entries of the unknown mmWave MIMO channel matrix in \cite{mmWave5}. Secondly, mmW/THz channels usually exhibit high-resolution angular-domain characteristics. Accordingly, beam tracking methods can be used to extract the CSI. In \cite{mmWave6}, a beam selection scheme was presented to decrease the complexity of beam tracking. Moreover, beam alignment was applied to improve the performance of beam tracking in \cite{mmWave7}.

\subsection{Massive Non-orthogonal Access}
A promising approach for increasing the number of supported access devices over a limited radio spectrum is the use of non-orthogonal multiple access (NOMA) techniques, which allow multiple devices to share the same time-frequency resource block. Hence, NOMA is a candidate technique for B5G wireless networks \cite{NOMA1}-\cite{JR:Zhiqiang_NOMA}. Compared to OMA techniques, NOMA techniques have the potential to improve spectral efficiency. In other words, for a given required spectral efficiency per device and a given bandwidth, NOMA can admit significantly more devices than OMA. Thereby, NOMA is able to support massive access in a limited radio spectrum. However, NOMA techniques lead to severe co-channel interference, especially in the massive access scenario. The key to realizing massive NOMA is interference management \cite{NOMA4}. So far, academia and industry have proposed several NOMA schemes, which can be classified into two categories, namely power-domain non-NOMA (PD-NOMA) and code-domain NOMA (CD-NOMA).


\begin{table*}\centering
\caption{Massive Non-Orthogonal Access Schemes.}
\label{Tab6}
\scriptsize
\begin{tabular}{|l|l|l|}\hline
 Reference & Type & Characteristics\\
\hline\hline X. Chen \emph{et al.} \cite{NOMA12} &  \tabincell{l} {PD-NOMA} & \tabincell{l} {Superposition coding at the transmitter and SIC at the receiver} \\
\hline Y. Du \emph{et al.} \cite{NOMA16} & \tabincell{l} {LDS-CDMA} & \tabincell{l} {Spreading of the transmitted symbols in the time domain by a low-density code and MAP at the receiver} \\
\hline R. Razavi \emph{et al.} \cite{NOMA17} & \tabincell{l} {LDS-OFDM} & \tabincell{l} {Spreading of the transmitted symbols in the frequency domain by a low-density code and MAP at the receiver} \\
\hline Z. Yuan \emph{et al.} \cite{NOMA18} & \tabincell{l} {MUSA} & \tabincell{l} {Spreading of the transmitted symbols by a code selected from a set of multiple sparse codes and MAP at the receiver}\\
\hline F. Wei \emph{et al.} \cite{NOMA21} & \tabincell{l} {SCMA} & \tabincell{l} {Mapping of the transmitted symbols into a codeword of a codebook consisting of multiple sparse codes and AMP at the receiver}\\
\hline
\end{tabular}
\end{table*}

\subsubsection{Power-Domain Non-Orthogonal Multiple Access}
PD-NOMA shares the radio spectrum through superposition coding with the transmit powers as weight factors \cite{NOMA5}. In this case, the access devices can be separated in the power domain. In order to decrease the co-channel interference caused by non-orthogonal transmission, successive interference cancelation (SIC) is usually carried out at the receiver. Specifically, the receiver first decodes the interfering signal with the highest transmit power and removes it from the received signal. Then, it decodes the interfering signal with the next highest transmit power until the desired signal is recovered. Hence, power allocation has a great impact on the performance of PD-NOMA. Intuitively, a device with a small channel gain is allocated a high transmit power, so as to guarantee fairness. Yet, it is not a trivial task to perform optimal power allocation in PD-NOMA, since there is residual inter-user interference from the devices with lower transmit powers. In \cite{NOMA6}, a cognitive power allocation scheme was proposed for two-user PD-NOMA. Since SIC only cancels the partial co-channel interference caused by the devices with higher transmit powers, devices with small channel gains may still suffer from strong co-channel interference after SIC, resulting in poor performance. In order to guarantee fairness, the authors in \cite{NOMA7} proposed a power allocation scheme that maximizes the rate of the device with the smallest channel gain.

PD-NOMA improves spectral efficiency at the cost of high computational complexity due to the use of a SIC receiver for interference mitigation. The computational complexity increases as the number of devices increases. In the scenario of massive access, the computational complexity and the signal processing delay might be prohibitive if SIC is performed for all devices. A possible solution to overcome these challenges is to perform device clustering \cite{NOMA8}, \cite{NOMA9}. Thereby, the devices are grouped into several clusters, where each cluster contains a small number of devices. SIC is performed within each clusters, which reduces the computational complexity effectively. In \cite{NOMA10}, a frequency-domain clustering scheme was proposed, where two devices assigned to the same subcarrier of an OFDMA system form a cluster. Frequency-domain clustering guarantees orthogonality across clusters, but decreases the spectral efficiency. Considering that the BTSs of B5G wireless networks will be equipped with a large-scale antenna array, it may be preferable to perform device clustering in the spatial domain, where each cluster can occupy the entire spectrum \cite{NOMA11}. Since spatial device clustering incurs extra inter-cluster interference, spatial beamforming combined with power allocation has to be employed to combat the interference \cite{NOMA110}. To further unleash the potential of non-orthogonal signaling, a fully non-orthogonal access scheme was designed for massive access in \cite{NOMA12}, where non-orthogonal pilot sequences were used to estimate the CSI with small overhead, and the estimated CSI was applied for the design of spatial beamforming for interference cancellation. As mentioned earlier, most IoT devices are simple nodes with limited computational capability. As a result, IoT devices may perform SIC imperfectly, resulting in severe residual intra-cluster interference. In this context, the authors in \cite{NOMA13} proposed a spatial beamforming scheme for massive access taking imperfect SIC into consideration. In fact, the design of a large number of spatial beams canceling inter-cluster interference also entails high complexity in the massive access scenario. In order to reduce the computational complexity of beam design, a beamspace non-orthogonal multiple access scheme was proposed in \cite{NOMA14}, which constructed the transmit beams based on statistical CSI. Compared to the beamforming schemes based on instantaneous CSI, the one based on statistical CSI leads to a complexity reduction at the cost of a loss in performance.

\subsubsection{Code-Domain Non-Orthogonal Multiple Access}
CD-NOMA assigns different codes to devices for multiplexing \cite{NOMA15}. Different from conventional CDMA, the assigned codes are sparse. However, the sparse codes can still offer spreading gains for suppressing undesired co-channel interference. Hence, only a simple message passing algorithm (MPA) at the receivers is needed to detect the sparse CD-NOMA sequences. Low-density signature CDMA (LDS-CDMA) \cite{NOMA16} and low-density spreading OFDM (LDS-OFDM) \cite{NOMA17} are two direct extensions of CD-NOMA. In particular, in LDS-CDMA, the symbol to be transmitted is spread in the time domain, while in LDS-OFDM, the chips are transmitted in the frequency domain. LDS-CDMA and LDS-OFDM can be selected based on the massive access system requirements.

In practice, LDS-CDMA and LDS-OFDM spread the signal using a predetermined sparse code for each device. In fact, if the device has multiple spreading codes, it is possible to further improve the performance of massive access. Inspired by this idea, multi-user shared access (MUSA) was proposed. For MUSA, there is a set of spreading codes \cite{NOMA18}. Each device randomly selects a spreading code for each symbol, and thus in fading channels, the average interference is suppressed due to the use of different spreading codes at the cost of a high-complexity receiver.

Unlike the above CD-NOMA schemes, sparse code multiple access (SCMA) maps the symbol to a sparse code \cite{NOMA19}. Each device has a predetermined codebook containing multiple sparse codes, where the nonzero elements are in the same positions. The symbol to be transmitted is mapped to an index, and the corresponding sparse code in the codebook is selected for transmission. The codebook design for SCMA was discussed in \cite{NOMA20} to further improve the performance of SCMA. Considering the requirements of massive access, an SCMA scheme with joint channel estimation and data decoding was proposed in \cite{NOMA21}.

A comparison of different massive non-orthogonal access schemes is provided in Table \ref{Tab6}. In summary, both PD-NOMA and CD-NOMA exploit new degrees of freedom for channel sharing so as to support massive access over limited wireless resources. However, both massive access techniques require sophisticated transceivers to combat co-channel interference. Considering that the channel matrices in massive access are high dimensional due to the deployment of the large-scale antenna arrays at the BTSs, the computational complexity at the transceivers may be prohibitive. Therefore, the design of simple but effective transceivers is an important topic for future research.

\section{Massive Coverage Enhancement}
IoT has found various applications in industry, agriculture, traffic, medicine, etc. Hence, IoT devices are distributed over a very large range, not only in urban areas, but also in rural areas. To decrease the power consumption and achieve long battery usage periods, e.g. 20 years, the transmit power of IoT devices is typically smaller than $23$ dBm. Therefore, the signal received from cell-edge devices is usually very weak, such that it is difficult to satisfy the QoS requirements. Consequently, the coverage of current cellular IoT is limited. Especially, signals received from indoor wireless devices are usually weak, but there is a large number of such devices. As a result, indoor massive access is a critical issue. 5G NB-IoT adopts several coverage enhancement schemes, e.g., low-order modulation and retransmission, to improve the quality of signals originating from the cell-edge and indoors. These schemes enhance the coverage at the cost of a low resource utilization efficiency. Yet, for massive access, there is no extra resource that can be used for coverage enhancement. Moreover, current cellular networks only cover densely populated areas, but IoT has been applied also in rural areas. Deploying new cellular networks in rural areas is inefficient in terms of capital cost. Therefore, it is necessary to develop new coverage enhancement strategies for massive access in B5G wireless networks. In the following, we discuss three possible coverage enhancement strategies.

\subsection{Cell-Free Massive MIMO}
\begin{figure}[t]
\centering
\includegraphics [width=0.45\textwidth] {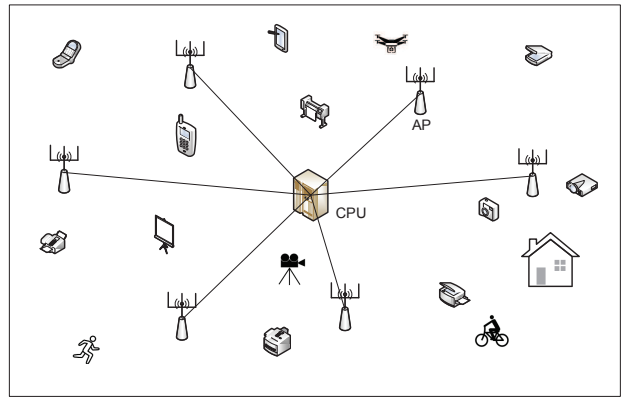}
\caption {A cell-free massive MIMO system, where multiple access points distributed over the whole area connect to a central processing unit through high-capacity optical fibre links. Thus, the access distances are shortened significantly.}
\label{Fig6}
\end{figure}

Massive MIMO has been proved to be an effective approach to enhance the coverage by exploiting its large array gain \cite{CE0}. Recently, it has been shown that cell-free massive MIMO can further improve coverage performance. Cell-free massive MIMO is indeed a distributed antenna system, which comprises a large number of access points (AP) connected to a central processing unit (CPU) \cite{CE1}, \cite{CE2}, as shown in Fig. \ref{Fig6}. Each AP can deploy one or multiple antennas. The system is not partitioned into cells and each user is served by one or multiple APs. Compared to co-located massive MIMO, cell-free massive MIMO significantly shortens the access distance, and thus broadens the coverage area. Since the APs form a large-scale antenna array, the same high spectral efficiency as with conventional massive MIMO can be achieved. In \cite{CE3}, it was proved that under uncorrelated shadow fading conditions, cell-free massive MIMO provides a nearly five-fold improvement in the 95\%-likely per-user throughput over a small-cell architecture, which is an enhancement strategy for 4G LTE \cite{Smallcell}, and a ten-fold improvement under correlated shadow fading conditions. Thus, cell-free massive MIMO is a promising choice for outdoor coverage enhancement at low power.

In practice, the APs equipped with independent RF chains are connected to the CPU by high-capacity optical fibres. Since the devices are randomly distributed over the service area, each AP has a different impact on the overall performance. Hence, it is necessary to wisely allocate the wireless resources to the APs to achieve the optimal system performance. A max-min power control scheme was proposed in \cite{CE4} to provide equal throughput for all users. It was found that most APs transmitted at less than the maximum possible power. Moreover, to improve the utilization efficiency of the low-resolution ADCs in cell-free massive MIMO, a quantization bit allocation scheme was proposed in \cite{CE5}.

\subsection{Intelligent Reflecting Surface}
\begin{figure}[t]
\centering
\includegraphics [width=0.45\textwidth] {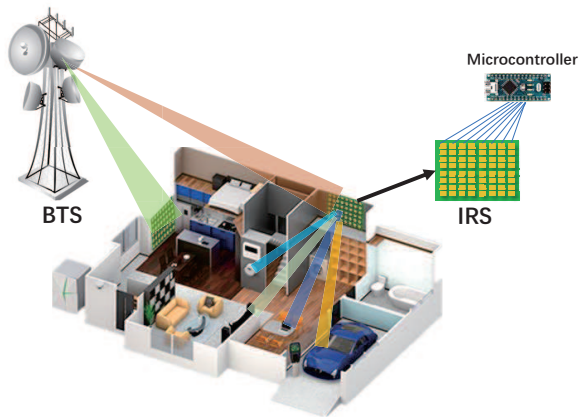}
\caption {Indoor coverage enhancement with an intelligent reflecting surface, where the intelligent reflecting surface comprising a large number of reflecting units generates a favorable propagation environment via beamforming and is controlled by a microcontroller.}
\label{Fig7}
\end{figure}

The deployment of IoT devices located indoors is expected to increase significantly in the coming decade \cite{CE06}. For indoor applications, in general, the received signal is weak due to the attenuation caused by walls \cite{CE6}. Hence, indoor coverage enhancement is crucial for realizing massive access in indoor scenarios. Considering the difficult propagation environment, indoor coverage enhancement is not a trivial task. A possible solution to this problem is to control the reflection characteristics of walls to establish favorable signal propagation environments. The concept of an intelligent wall as an autonomous part of a smart indoor environment was proposed in \cite{CE6}. In particular, an intelligent wall is a wall equipped with an active frequency-selective surface, simple low-cost sensors, and a cognitive engine, which can control the radio coverage to improve the overall system performance. A simple but effective way to realize such a wall is the deployment of a reconfigurable reflect-array, also referred to as an intelligent reflecting surface (IRS), with a large number of reflecting units that reflect the transmitted signals \cite{CE7}, as shown in Fig. \ref{Fig7}.

By optimally controlling the phase shift of each unit of the IRS, the desired signal can be enhanced while undesired interference can be canceled \cite{CE70}, \cite{CE71}. Since the IRS comprises a large number of reflecting units, it can play the same role as a large-scale antenna array through spatial beamforming. As a result, even for a large number of indoor wireless devices, the quality of the received signals can be significantly improved. However, compared to the large-scale antenna array, the IRS entails lower cost and complexity. This is because the large IRS does not require power-hungry RF chains. In other words, IRS is a green massive coverage enhancement strategy. In \cite{CE8}, the IRS phase shifters were optimized to maximize the sum rate in a multiuser scenario. Taking into account that the number of phase shifts for each reflecting unit is finite in practice, IRS beamforming was optimized to minimize the total power consumption in \cite{CE9}. It is worth pointing out that IRS can be utilized to enhance not only indoor coverage, but also outdoor coverage. Hence, IRS is regarded as a promising technique for B5G wireless networks \cite{CE10}-\cite{CE110}. Exploiting IRS specifically for massive access is an interesting topic for future work.

\subsection{Satellite Communications}
The rural deployment of IoT is important for monitoring and management applications \cite{CE120}. For instance, security cameras have been installed in forests to predict wildfires \cite{CE121}, and a large number of sensors have been deployed in the sea to monitor the ocean resources \cite{CE122}. Currently, these areas are not covered by cellular networks. From a cost perspective, it is prohibitively expensive to deploy new cellular networks in rural areas. Thereby, satellite communications are expected to provide wireless access in rural areas \cite{CE123}. In particular, a satellite can cover a large area, and thus the cost of wireless access is substantially decreased. Hence, satellite communication is expected to become an important component of B5G wireless networks \cite{CE12}, \cite{CE13}. By integrating space and ground networks, seamless coverage can be provided all over the world.

\begin{figure}[ht]
\centering
\includegraphics [width=0.45\textwidth] {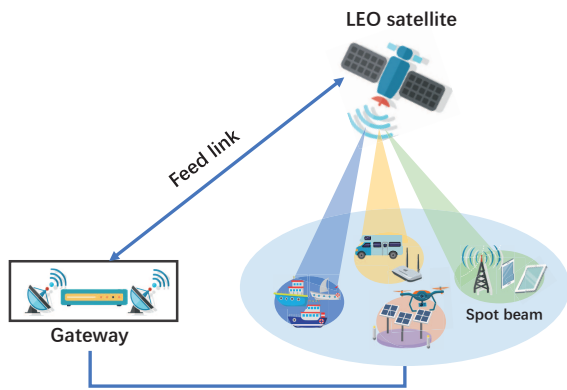}
\caption {Coverage enhancement by a multi-beam LEO satellite, which can provide low-latency and reliable wireless access to a large number of devices distributed in large rural areas by using multiple spatial spot beams.}
\label{Fig8}
\end{figure}

To shorten the access latency, low earth orbit (LEO) satellites are usually used as access points for space networks \cite{CE14}. By applying multiple-beam techniques, a LEO satellite can serve a large number of devices simultaneously, as shown in Fig. \ref{Fig8}. In \cite{CE15}, the transmit beamforming was designed for multicast in multiple-beam satellite communications. Considering imperfect CSI at the satellite, a robust beamfoming scheme was developed in \cite{CE16} with the objective of minimizing the total power consumption. Moreover, cooperative multicast transmission in integrated space-ground networks was investigated in \cite{CE17}. Furthermore, a max-min beamforming scheme was designed to jointly optimize the beamforming vectors of the BTS and the satellite. With the fast evolution of high-throughput multiple-beam satellite communications, satellite IoT has been proposed in \cite{CE18}, and is expected to accelerate the creation of an IoE.

Massive access is subject to complex and time-varying propagation environments. Hence, it is natural to enhance wireless coverage by combining multiple strategies. For example, in hotspot areas, both cell-free massive MIMO and IRS can be employed to enhance the signal quality. However, the combination of multiple enhancement strategies entails high implementation complexity and cost, which have to be carefully considered for practical deployment.

\section{Other Massive Access Topics}
As outlined in Sections III-VI, massive access in B5G wireless networks involves many aspects, including theoretical concepts, protocol design, algorithm development, and coverage extension, which have to be jointly considered to improve the efficiency and reliability of massive access. However, in order to realize massive access in B5G wireless networks, there are additional critical issues which have to be considered such as energy supply and access security. Herein, we provide a brief discussion of these two topics from the perspective of massive access.

\subsection{Wireless Energy Transfer}
Currently, most IoT devices are battery powered. Since the battery capacity of IoT devices is quite limited, the transmit power has to be very low, e.g., 23 dBm. The low transmit power limits the capabilities of IoT applications. On the other hand, for higher transmit power, the battery has to be replaced frequently. The battery replacement for a massive number of IoT devices entails a high human cost and a large environmental strain. Moreover, it is difficult to replace the batteries of devices in extreme environments, e.g., in walls and under water. Recently, wireless energy transfer, namely wireless charging, has received considerable attention from both academia and industry \cite{WPT1}-\cite{WPT3}. In particular, wireless energy transfer based on RF signals can provide stable and reliable energy supply. Hence, IoT devices can realize sustainable communications even under adverse conditions, as long as there is wireless coverage. More importantly, due to the broadcast nature of wireless channels, many devices can be charged in parallel. Hence, wireless energy transfer is particularly appealing for cellular IoT with massive access \cite{WPT4}, \cite{WPT5}.

A challenging issue in wireless energy transfer is the low energy transfer efficiency due to path loss and channel fading during the transmission of the wireless energy signal. As a result, the effective distance of wireless energy transfer is too short to achieve the broad coverage desirable for massive access. To overcome this problem, the concept of energy beamforming was introduced for wireless energy transfer \cite{WPT6,JR:energy_beamforming}. Specifically, by using spatial beamforming, the energy signal is focused on the receiver, and thus the transfer efficiency can be improved effectively. It was shown that even with partial CSI at the transmitter, energy beamforming can enhance the energy transfer efficiency. Especially in multiuser scenarios, energy beamforming can facilitate simultaneous energy harvesting for a massive number of devices. Furthermore, when the transmitter is equipped with a large-scale antenna array, the effective transmission distance can be increased significantly \cite{WPT7}. By exploiting the very high spatial resolution of large-scale antenna arrays, it is possible to charge a massive number of devices with high efficiency. Moreover, multiple-point cooperation and relaying can be employed to further increase the transfer distance \cite{WPT8}-\cite{WPT10}.


Wireless energy transfer is already being applied for short-distance charging scenarios. For instance, mobile phones can be charged without wireline connection. The provision of long-distance wireless energy transfer for practical massive access is still an open research problem due to the low transfer efficiency and requires further research.

\subsection{Physical-Layer Security}
In massive access, a massive number of devices share the radio spectrum. Any device can receive the other devices' signal, resulting in the risk of information leakage. Traditionally, access security has been realized by using upper-layer encryption techniques \cite{PLS1}, \cite{PLS2}. Due to the fast evolution of communication technology, the computational capabilities of eavesdropping nodes have significantly increased. Consequently, encryption techniques have to become more sophisticated to guarantee information security. Yet, most IoT devices are low-cost nodes with limited computational capability, and thus they cannot afford the high complexity required for advanced encryption techniques. Moreover, for some versions of massive access, e.g., grant-free random access, conventional encryption techniques relying on secure key distribution are not applicable. In this context, physical-layer security techniques, as a complement to conventional encryption techniques, can be adopted to facilitate secure massive access \cite{PLS3}. The essence of physical-layer security is to exploit the inherent random characteristics of wireless channels, e.g., fading, interference, and noise, to ensure that the information transmission rate of the desired link is higher than the eavesdropping channel capacity, and hence, the eavesdropper is not able to decode the intercepted signal correctly \cite{PLS30}, \cite{PLS300}. As mentioned earlier, massive access causes severe co-channel interference, which can be exploited to improve the security of B5G wireless networks by applying physical-layer security techniques.

According to the basic principles of physical-layer security, in order to enhance the secrecy performance, it is necessary to improve the quality of the legitimate signal and decrease the quality of the eavesdropping signal simultaneously. Hence, multiple-antenna techniques are commonly employed to provide physical-layer security \cite{PLS4}. For instance, if the legitimate signal is transmitted in the null space of the eavesdropping channel matrix, the eavesdropper cannot receive the legitimate signal. More generally, it is possible to maximize the secrecy rate through spatial beamforming. Even in challenging environments with multiple eavesdroppers, spatial beamforming can facilitate access security if there are enough spatial degrees of freedom at the transmitter \cite{PLS5}. Unfortunately, the secrecy performance of spatial beamforming heavily depends on the accuracy of the CSI available at the multiple-antenna transmitter. In general, the CSI of the eavesdropping channel is difficult to obtain, since eavesdroppers usually hide their existence by remaining silent (i.e., passive eavesdroppers). In this case, artificial noise may be sent in the null space of the legitimate devices' channel matrices to confuse the eavesdroppers \cite{PLS6}.

In B5G wireless networks, the BTSs might be equipped with a large-scale antenna array. By exploiting the very high spatial resolution of the large-scale antenna array, secure access for a massive number of devices can be provided. It has been proved that if the BTS has full CSI, linear precoding can ensure that the information leakage asymptotically tends to zero \cite{PLS7}. Hence, even without the eavesdroppers' CSI, it is possible to realize secure massive access. However, the acquisition of the legitimate devices' CSI in massive MIMO systems with a large number of devices is not trivial \cite{PLS8}. Firstly, pilot sequences are usually non-orthogonal, resulting in low CSI accuracy. Secondly, the eavesdroppers can send interfering signals during channel estimation to increase the interception probability. Hence, providing physical-layer security in massive access is still a challenging issue.

\section{Future Research Directions}
Despite the significant research efforts dedicated to facilitating massive access in B5G wireless networks, many challenging issues remain to be tackled. In the following, we discuss some future research directions.

\subsection{Mobile Access}
In cellular IoT, a fraction of the devices is expected to be mobile and some devices may move with high speed. Mobility gives rise to additional challenges for massive access. First, mobility leads to fast time-varying channel fading making the acquisition of the accurate CSI needed to facilitate massive access very challenging, resulting in a performance degradation. Second, mobility causes frequent handoffs. For example, in Internet-of-Vehicles applications, frequent handoffs between BTSs may occur. Hence, the priority of mobility handoffs and new access requests have to be properly handled. Moreover, mobility may change the channel capacity of massive access \cite{Mobility1}. So far, only a few works have considered mobility in massive access \cite{Mobility2}, \cite{Mobility3}.

\subsection{Modulation and Coding}
Modulation and coding schemes (MCS) are key for guaranteeing both high efficiency and high reliability for massive access. The 5G wireless standard utilizes low-density parity check (LDPC) codes and polar codes for the data and control channels, respectively \cite{Multipleaccess}. For massive access in B5G wireless networks, some new characteristics have to be considered. Firstly, the sporadic nature of IoT traffic favors the use of short packets, and thus, short FEC codes should be adopted. Secondly, as IoT devices are typically simple nodes with limited computational capabilities, MCS in B5G wireless networks have to be low-complexity. Therefore, the design of new low-complexity short codes for massive access is a key research problem.

\subsection{Big Data Analytics and Large Dimensional Signal Processing}
In B5G wireless networks, there is a massive number of IoT devices generating a huge volume of data. Meanwhile, since the BTS is usually equipped with a large-scale antenna array, the dimension of the received signal is very large. Hence, massive access inevitably leads to big data in volume and dimension. This significantly increases the burden on B5G wireless networks. In order to improve the efficiency of massive access, it is necessary to develop methods for big data analytics and large dimensional signal processing. For instance, a dimension reduction-based algorithm was designed to decrease the computational complexity of massive active device detection in B5G wireless networks \cite{Bigdata}. However, there is still a lack of efficient methods for channel estimation, precoding design, and other aspects of massive access. Developing such methods is an exciting future research direction.

\subsection{Ultra-Reliable Low-Latency Communication}
Ultra-reliable low-latency communication (URLLC) is a basic requirement in many cellular IoT application scenarios, e.g.  Internet-of-Vehicles \cite{URLLC1}. However, it is very challenging to guarantee URLLC over fading channels. First, massive access leads to severe co-channel interference, which decreases the access reliability. Second, short packets are used in cellular IoT to decrease the latency, but they are also prone to a high decoding error rate. Hence, achieving URLLC for massive access is still an open issue.

\subsection{Machine Learning-Based Massive Access}
Smart communication is a new trend in wireless communications. Currently, machine learning, especially deep learning, is being applied in wireless communications for resource allocation, signal processing, channel estimation, and transceiver design \cite{Deeplearning}. It has been shown that machine learning can decrease the design complexity of wireless communication networks while achieving high performance. In cellular IoT, the BTS has to cope with the wireless access of a massive number of devices, resulting in a high computational complexity. The application of machine learning for massive access is expected to significantly decrease complexity. Yet, there is a lack of analytical frameworks for machine learning as applied in wireless networks, which currently limits its applicability in practice \cite{Lizhong}.

\subsection{Convergence of Sensing, Computation, and Communication}
Sensing, computation, and communication are three basic functionalities of B5G wireless networks \cite{Integration1}. Traditionally, these three functionalities have been carried out independently. Hence, it is necessary to allocate wireless resources for each functionality, resulting in a high resource consumption. In the case of massive access, the resources required to support these functionalities might be prohibitive. Hence, it is desirable to jointly design these three functionalities to improve the efficiency of massive access. For instance, the transmission of sensed signals over multiple access channels can also be exploited to perform computations by applying over-the-air computation techniques in \cite{Integration2}. In this scenario, the limited wireless resources can be utilized with high efficiency, especially for massive access. Therefore, the convergence of sensing, computation, and communication is an important future direction for cellular IoT in B5G wireless networks.

\section{Conclusion}
This paper provided a comprehensive review of massive access in B5G wireless networks from different perspectives. First, we summarized the basic characteristics of massive access, such as low power, massive connectivity, and broad coverage. Then, we surveyed information theoretical concepts for massive access, focusing on massive random access and massive short-packet access. Next, we discussed massive access protocol design, with an emphasis on grant-free random access protocols. In particular, we presented the sensing matrix design and the corresponding device activity detection algorithms, including optimization algorithms, greedy algorithms, and Bayesian algorithms. Subsequently, we provided an overview of massive orthogonal and non-orthogonal access techniques, respectively. Furthermore, we identified challenges for massive coverage enhancement in outdoor, indoor, and rural environments. Finally, we discussed potential challenges for providing massive access in B5G wireless networks and pointed out some possible future research directions.

\end{document}